\documentclass[aps,prl,reprint,superscriptaddress,nofootinbib]{revtex4-1}

\usepackage{lineno}
\usepackage{graphicx}
\usepackage{float}
\usepackage{amsmath}
\usepackage{amssymb}
\usepackage{tikz}
\usepackage{siunitx}
\usepackage{overpic}
\usepackage{xcolor}

\makeatletter
\renewcommand{\fnum@figure}{\textbf{Fig. \thefigure}}
\makeatother

\begin{document}

\title{Additively manufactured ultra-high vacuum chamber below $10^{-10}$ mbar}



\author{N. Cooper}
\email[]{nathan.cooper@nottingham.ac.uk}
\affiliation{\mbox{School of Physics and Astronomy, University of Nottingham, Nottingham, NG7 2RD, UK}}
\author{L. A. Coles}
\email[]{laurence.coles@addedscientific.com}
\affiliation{\mbox{Added Scientific Ltd., Isaac Newton Centre, Nottingham Science Park,  Nottingham, NG7 2RH, UK}}
\author{S. Everton}
\affiliation{\mbox{Added Scientific Ltd., Isaac Newton Centre, Nottingham Science Park, Nottingham, NG7 2RH, UK}}
\author{R. P. Campion}
\affiliation{\mbox{School of Physics and Astronomy, University of Nottingham, Nottingham, NG7 2RD, UK}}
\author{S. Madkhaly}
\affiliation{\mbox{School of Physics and Astronomy, University of Nottingham, Nottingham, NG7 2RD, UK}}
\author{C. Morley}
\affiliation{\mbox{School of Physics and Astronomy, University of Nottingham, Nottingham, NG7 2RD, UK}}
\author{J. O'Shea}
\affiliation{\mbox{School of Physics and Astronomy, University of Nottingham, Nottingham, NG7 2RD, UK}}
\author{W. Evans}
\affiliation{\mbox{Department of Physics and Astronomy, University of Sussex, Brighton, BN1 9QH, UK}}
\author{R. Saint}
\affiliation{\mbox{Department of Physics and Astronomy, University of Sussex, Brighton, BN1 9QH, UK}}
\author{P. Kr{\"u}ger}
\affiliation{\mbox{Department of Physics and Astronomy, University of Sussex, Brighton, BN1 9QH, UK}}
\author{F. Oru\v{c}evi\'{c}}
\affiliation{\mbox{Department of Physics and Astronomy, University of Sussex, Brighton, BN1 9QH, UK}}
\author{C. Tuck}
\affiliation{\mbox{Faculty of Engineering, University of Nottingham, Nottingham, NG7 2RD, UK}}
\affiliation{\mbox{Added Scientific Ltd., Isaac Newton Centre, Nottingham Science Park, Nottingham, NG7 2RH, UK}}
\author{R. D. Wildman}
\affiliation{\mbox{Faculty of Engineering, University of Nottingham, Nottingham, NG7 2RD, UK}}
\affiliation{\mbox{Added Scientific Ltd., Isaac Newton Centre, Nottingham Science Park, Nottingham, NG7 2RH, UK}}
\author{T. M. Fromhold}
\affiliation{\mbox{School of Physics and Astronomy, University of Nottingham, Nottingham, NG7 2RD, UK}}
\author{L. Hackerm{\"u}ller}
\affiliation{\mbox{School of Physics and Astronomy, University of Nottingham, Nottingham, NG7 2RD, UK}}

%



\begin{abstract}
We demonstrate the first additively manufactured ultra-high vacuum (UHV) chamber, and show that it reaches a pressure below 10$^{-10}$\,mbar. The chamber's mass is less than one third of that of industry standard equivalents. Without powered pumping it can operate for 48 hours in the 10$^{-9}$ mbar regime. Analysis of the material surface reveals that surface oxides, generated during the manufacturing process, play a key role in enhancing vacuum performance. 
We use the chamber to trap a cloud of cold $^{85}$Rb atoms – the starting point for many precision timekeeping and sensing devices – and so demonstrate its suitability for quantum technologies. Our findings enable the creation of lightweight, performance-optimized, compact UHV systems as well as functionalized materials and surfaces created by additive manufacturing for UHV environments. Our results thus have the potential to revolutionize all areas relying on UHV. 
\end{abstract}

\maketitle


\section*{Introduction}
In this work, we show that ultra-high vacuum (UHV) compatible systems are achievable via additive manufacturing (AM), opening up opportunities for weight and design critical applications, including compactness and enhanced functionality across all sectors relying on high and ultra-high vacuum equipment.

Metal-based additive manufacturing, also known as 3D printing, represents a paradigm change in engineering and production methods, but so far could not be applied to ultra-high vacuum systems. AM methods allow components to be designed for optimal performance, without the constraint of traditional manufacturing considerations. 
This approach, known as design for AM \cite{Gibson2015}, is ideal for the production of custom parts and complex geometries, and enables advanced features such as mass reduction and enhanced stability via latticing, elimination of extraneous material, and part consolidation \cite{Gibson2015,Ghidini2018}, which reduces the number of separate components, lowering the risk of vacuum leaks and leading to highly compact systems. These techniques can yield compact mono-shell systems with topologies optimised for structural and electromagnetic performance.

Extending additive manufacturing techniques to UHV systems provides a step change in the design and functionality of equipment for multiple fields of science and technology including e.g. photosensors, cameras, cryostats and x-ray photo-electron spectroscopy analyzers, as well as the burgeoning field of quantum technologies.  Mass reduction and the creation of compact systems is of significant importance for e.g. portable and space-borne applications of quantum systems requiring UHV \cite{Ghidini2018,Becker2018}. Applying AM methods to UHV apparatus yields unprecedented reductions in the size, weight and development time and thus facilitates both fundamental research and technological development across a range of disciplines. 


Despite recent advances in AM processing of metals \cite{Martin2018,Zhu2018,Barriobero2018,Park2019}, printed UHV chambers have so far remained elusive and have even been thought to be impossible.  A reason for this lies in the generally rough surfaces, formation of pores \cite{Martin2019,Everton2016} and limited hardness of the materials resulting from AM processes \cite{Martin2018}. We show that these obstacles can be overcome using a suitable alloy and post-processing methods leading to tightly-connected, non-porous surfaces covered by a protective oxide layer. Surface analysis using mass spectrometry, scanning electron microscopy (SEM) and x-ray photoelectron spectroscopy (XPS) of the AM material unveils the critical and necessary role of the surface chemistry in improving UHV performance.

%
%
\begin{figure*}[!t]
\includegraphics[width=1.8 \columnwidth]{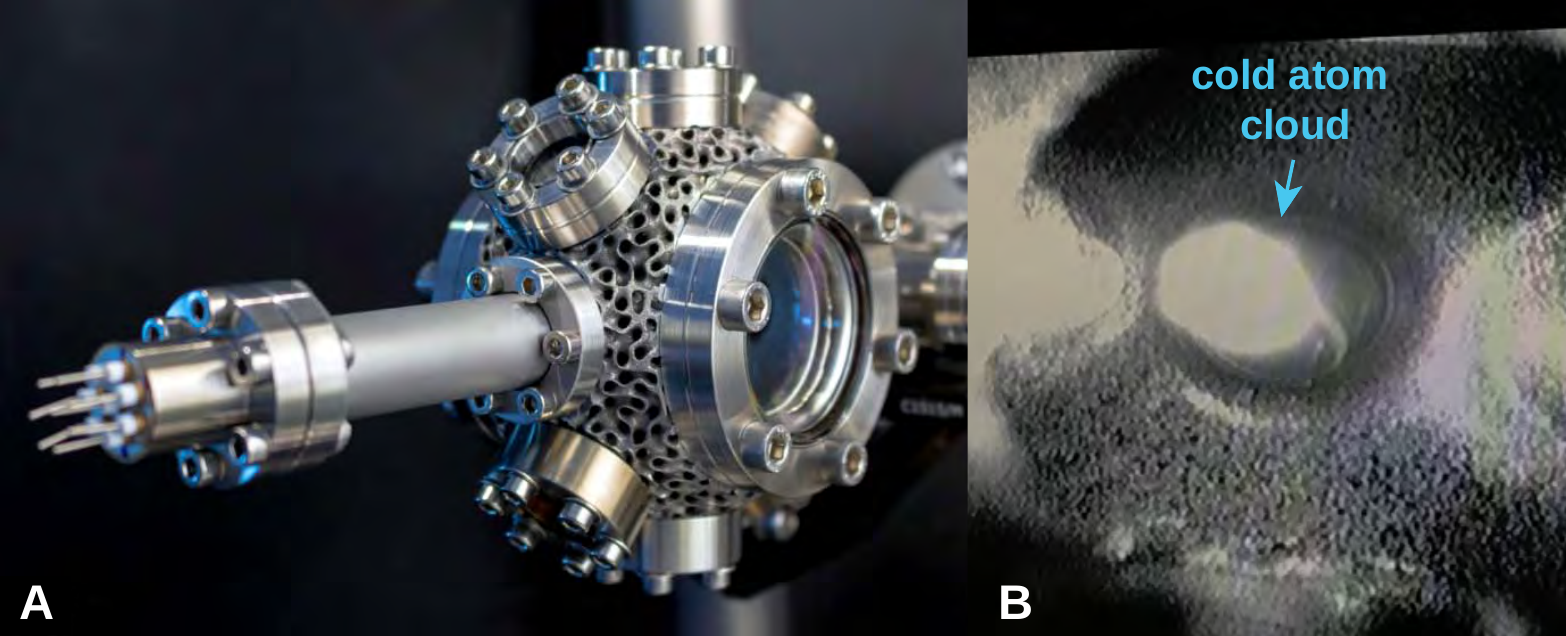} 
\caption{\textbf{An additively manufactured UHV chamber for quantum technologies.} (\textbf{A}) Photograph of the final, assembled AM-built UHV chamber with standard UHV components. (\textbf{B}) Fluorescence image of a cold cloud of $^{85}$Rb atoms confined in a magneto-optical trap inside the chamber.}
\end{figure*}

An important application example is quantum systems based on cold atoms, which are undergoing a rapid transition from lab-based research experiments to devices with wide scientific and commercial viability \cite{Knight2016}.  Portable cold atom systems are expected to underpin the next generation of sensing and timekeeping technologies \cite{Bongs2016, Knight2016}. Examples include atomic clocks for precision timekeeping \cite{Falke2014,Ludlow2015,Riehle2017}, high-precision gravimeters and gravi-gradiometers for applications in geology, navigation and civil engineering \cite{Bidel2018,Bin2014,Menoret2018} and cold atom magnetometers \cite{McGuirk2002,Behbood2013} with applications in medical imaging, navigation and archaeology. For all of these, reducing the mass of the vacuum equipment and enhancing its stability is essential. Low-mass UHV systems are also required for novel space-based sensors for fundamental research \cite{Dimopoulos2009,Carraz2014}, and proposed space-borne experiments \cite{Dimopoulos2009,Carraz2014, Liu2018}. 

Although there are a few examples of AM materials being subject to a UHV environment \cite{Saint2018,Vovrosh2018}, until now, no AM built system has been able to withstand the conditions necessary for performance at UHV. Previously achieved pressures were limited to $>10^{-5}$\,mbar \cite{Jenzer2017}.

The AM chamber demonstrated here is produced via a metallic powder bed fusion (M-PBF) technique \cite{Gibson2015} from aluminum alloy AlSi10Mg, with a mass of 245\,g, corresponding to a mass reduction of approximately 70\,\% compared to conventional stainless steel chambers of similar size. This reduction is an example of what can be achieved by applying AM methods. We demonstrate the operation of such a chamber for the creation of a cold atom cloud, the starting point for portable quantum sensors or space-based fundamental quantum experiments.

\begin{figure*} [!ht]   
\includegraphics[width=1.7\columnwidth]{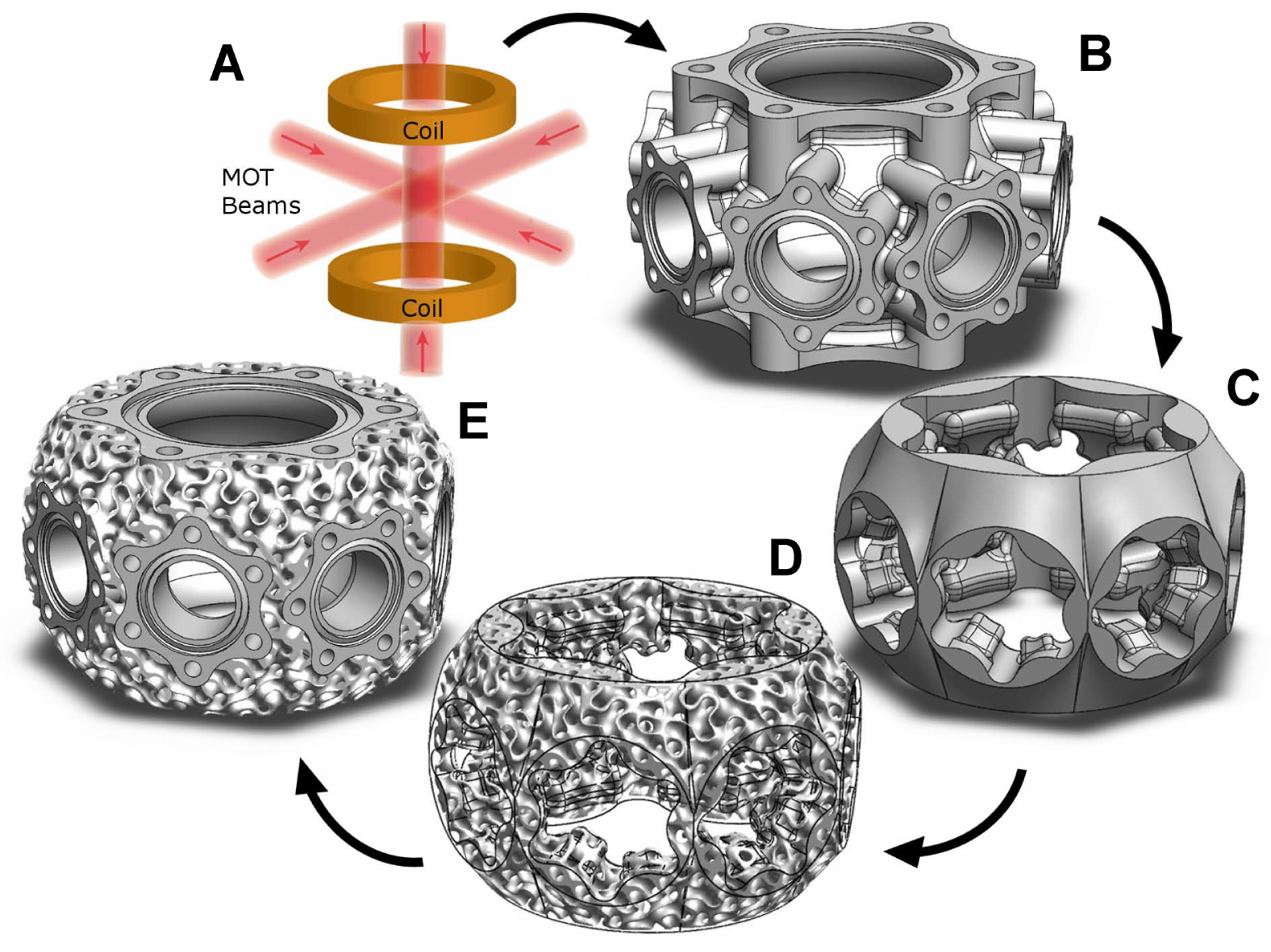} 
\caption{\textbf{Overview of the design process.} (\textbf{A}) Illustration of the functional requirements for cold atom cloud trapping. (\textbf{B}) Thin wall, symmetrical internal chamber design with conflat ports. (\textbf{C}) Volume identified for additional lattice support. (\textbf{D}) The variable density triply periodic minimal surface gyroid matrix lattice, which has been generated from the previous volume in (\textbf{C}) to add a layer of robustness and increase structural stiffness. (\textbf{E}) Final weight-optimized chamber design.}
\end{figure*}

\section*{Results}
\noindent \textbf{Design} 

\noindent The starting point for the vacuum chamber design is solely the functional requirements of a magneto-optical trap \cite{Metcalf1999} (Fig. 2A). These are: optical access for three orthogonal pairs of counter-propagating laser beams, magnetic coils, a UHV environment and the ability to connect to standard components, such as vacuum pumps. 

A “design for M-PBF” methodology was applied to tailor the chamber for this application, reducing the overall weight without affecting the mechanical stability. The light-weighting of the chamber core was achieved via refinement of the port geometry and minimizing the spacing between ports. A 2.5 mm thick internal skin was added between the ports to hold the UHV, as shown in Fig. 2B. For stability reasons, effort was made to keep symmetry where possible in the chamber design. The resulting (negligible) deformation of the chamber under an external pressure load was analyzed via ‘Finite Element Analysis’ (FEA) based simulation using ‘MSc Marc 2017.1.0’, see fig. S2.

Within a predefined volume (as shown in Fig. 2C) a variable density triply periodic minimal surface lattice was integrated with the symmetric internal core design to add a layer of robustness and increase stiffness; this took the form of a matrix based gyroid surface \cite{Maskery2018}, as shown in Fig. 2D. The variable volume fraction (density) of the lattice permits reinforcement of areas which need more support, such as those adjacent to the ports. It also provides the added benefit of passive cooling of the chamber due to the increased external surface area. The final chamber design (Fig. 2E) consists of multiple CF ports ($2\times$CF40 ports and $8\times$CF16) making it compatible with standard UHV equipment, as shown in Fig. 1A and Fig. 4A.

Of the metallic alloys available for AM processing, aluminum alloy AlSi10Mg was selected for this application due to its high specific strength and low density \cite{Chen2017}. Materials produced by M-PBF have a characteristically ultra-fine grain structure with epitaxial grain growth aligned with the build direction \cite{Thijs2013}. Solution heat treatments can be applied to homogenize the microstructure or control the grain size. These methods are often utilized to increase the ductility of the alloy; subsequently an ageing heat treatment is applied to promote precipitate hardening and increase the material strength \cite{Aboulkhair2015,Aboulkhair2016}. The resulting material performance after the selected heat treatment regime was verified for knife edge seal repeatability and thread torque requirements. The Vickers hardness was measured for a machined surface (as in Fig. 3G) as $(105\pm 0.8)$ HV5. See Materials and Methods for full details of post-processing and material performance tests. 

\vspace{3mm}

\noindent \textbf{Outgassing behavior and surface material analysis}

\noindent M-PBF components and materials tend to have rough surfaces \cite{Martin2018} and have therefore been thought unsuitable for UHV applications. We analyze the surface resulting from the printing and post-processing - surprisingly the surface roughness does not limit the suitability for UHV. Fig. 3A shows a focus variation microscopy (FVM) image of the surface of a test sample, built and post-processed in the same way as the UHV chamber. The surface roughness $Sq$, defined as the root mean square height over the area sampled, was measured as $Sq = 5\,\mu$m. Scanning Electron Microscopy (SEM) of the same surface reveals a rich surface structure with lateral features as small as 1-10$\,\mu$m (Fig. 3, B and C), but, importantly, no evidence for cracks, tears or deep pores. The dense, high-quality structure of the material is also demonstrated in SEM images of machined surfaces. Fig. 3D shows an image of the machined knife-edge, with some machining marks but no material defects visible at this zoom level. An image taken at higher magnification (Fig. 3E) shows small, micrometer-sized, defects aligned with the direction of the machining tool, but none orthogonal to it. This anisotropy suggests that the defects stem from the machining rather than the AM build process. Importantly, the defects seen on the machined surface are isolated and discontinuous, and no evidence of lateral cracking is observed. These defects would therefore not be expected to affect the vacuum performance of the chamber, as confirmed by the vacuum testing results described below.

The outgassing behavior was characterized via a mass spectrometric study on the M-PBF alloy used to build the vacuum chamber, while varying the temperature from 20 to $500^\circ$C. In addition, x-ray photo-electron spectroscopy (XPS) of a machined and unmachined sample surface was performed at room temperature. 

\begin{figure*}[!t]
\includegraphics[width=2.0\columnwidth]{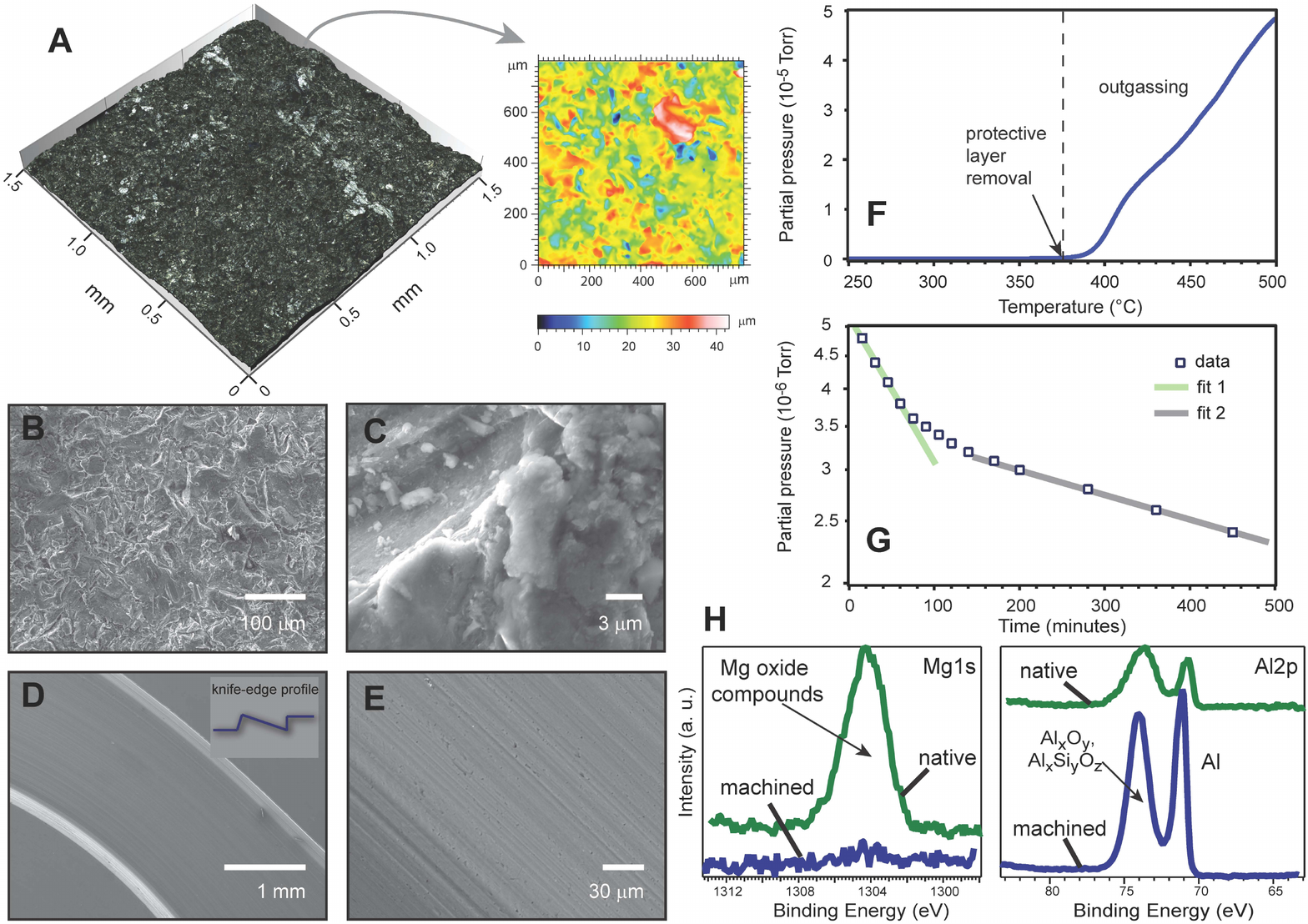}  
\caption{\textbf{Surface and material characterization.} (\textbf{A}) Focus Variation Microscopy image of the untreated surface of an additively manufactured sample of AlSi10Mg (left) and color map showing the corresponding surface profile (right). (\textbf{B,C}) SEM images of the untreated surface vs. (\textbf{D,E}) SEM images of the machined surface and knife-edge. (\textbf{F}) Material characterization with mass spectrometry - partial pressure of $^{24}$Mg in the Knudsen Cell as a sample of AM AlSi10Mg is heated to $500^\circ$C at $10^\circ$C per minute. (\textbf{G}) Partial pressure of $^{24}$Mg in the Knudsen Cell as a function of time after a sample of AM AlSi10Mg has been heated rapidly to 420$^\circ$C and held at this temperature. Lines are linear fits at short and long timescales. (\textbf{H}) XPS spectra showing a high content of Mg oxides in the untreated surface (left) vs. a low abundance of Mg oxides in the machined surface (right). }
\label{analysis}
\end{figure*}

%
The results of both measurements suggest the formation of an Mg-enriched, oxidized layer on the surface of the material (see discussion below), which plays a role in the suppression of outgassing. 

Samples of the M-PBF material ($10\times10\times4$\,mm cuboids), were placed in a temperature-controlled Knudsen Cell connected to a line-of-sight mass spectrometer. The molecular/atomic beam from the source can be interrupted by a rotating chopper, which allows the line of sight beam to be distinguished from background signals \cite{Campion2010}.  Scans were run across the mass range of 1 to 200\,amu, while the material was heated up as far as $500^\circ$C with a heating rate of $10^\circ$C/min (Fig. 3F). The experiment was then repeated in a quasi-stable situation, holding the sample for 48 hours each at temperatures of $250^\circ$C, $350^\circ$C, $400^\circ$C and $450^\circ$C in turn.  In both experimental settings, i.e. the rapid heating and the quasi-stable situation, similar behavior was observed. Other than (expected and unavoidable) atmospheric species, the only peaks detected were for magnesium (primarily $^{24}$Mg), which became evident above $400^\circ$C. Below $375^\circ$C, no $^{24}$Mg signal is measurable, as determined by the noise level, while above that temperature a sharp rise in the magnesium signal is visible. 

If the temperature is raised above $400^\circ$C and then subsequently decreased, the $^{24}$Mg signal does not return to its original value, but remains at a considerably higher level (factor $10^3$ at $350^\circ$C). The signal then persists  at temperatures far below $350^\circ$C. We interpret these results as the removal of a protective layer. 
The presence of this layer suppresses the rate of Mg emission from the sample, and may have a similar effect on the outgassing of other particle species. 

This interpretation is supported by a second experiment, where samples were held at $420^\circ$C for 150 minutes and the time variation of the emission was measured (Fig. 3G). All samples show a rapid initial reduction in Mg emission (fit 1 in Fig. 3G with a gradient of $g_{1}=-5.1(4)\exp(-3)\,$ln(Torr)/min) followed by a much slower subsequent tail-off (fit 2 with a gradient of $g_{2}=-9.1(2)\exp(-4)\,$ln(Torr)/min). 
For Fig. 3(F and G) the system was calibrated with a pure magnesium sample of the same dimensions as the M-PBF material and the observed count rate for the M-PBF samples was normalized to the magnesium counts and expressed as a partial pressure (fig. S3).

\begin{figure*}[!ht]
\includegraphics[width=1.8\columnwidth]{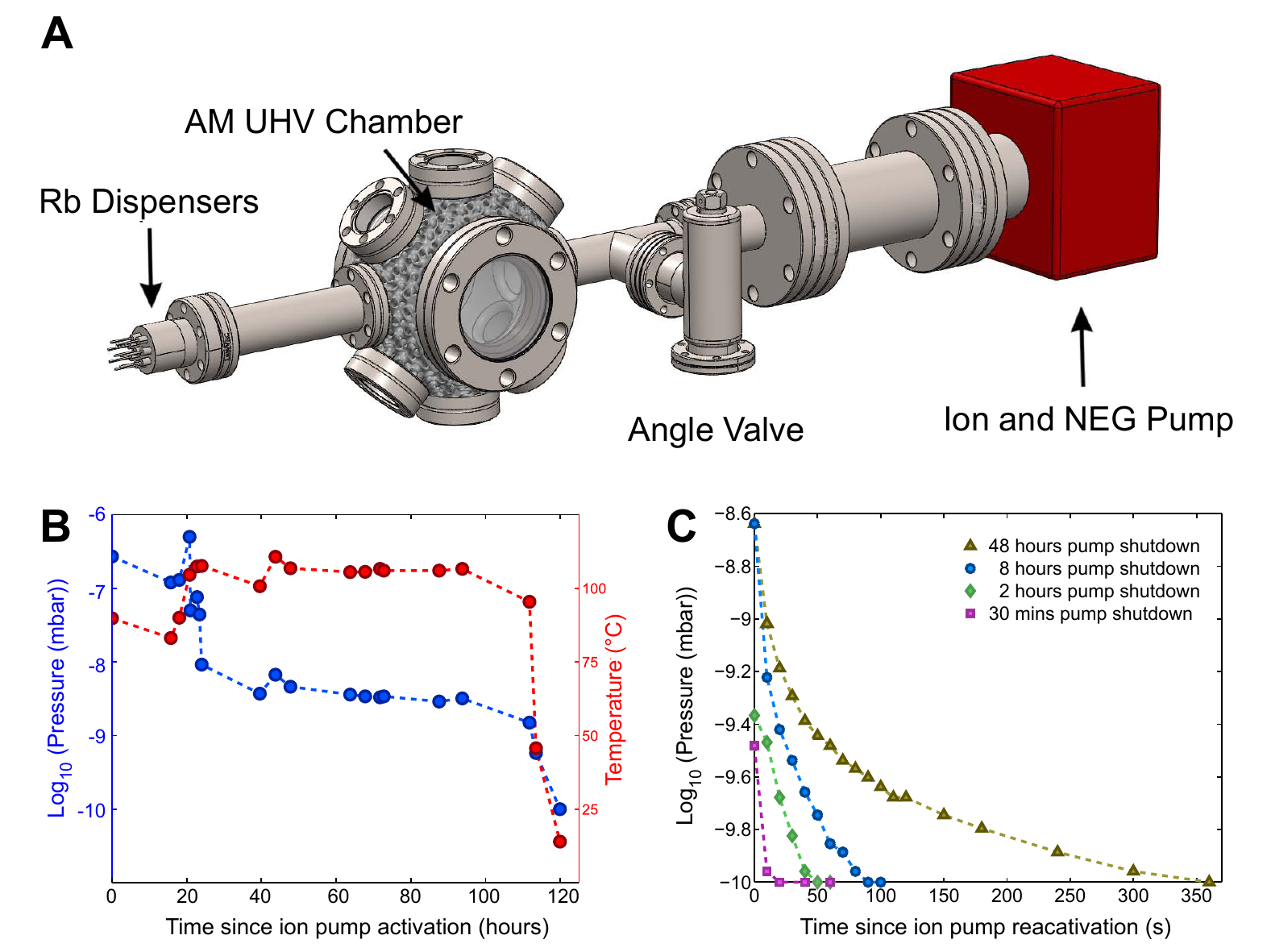}  
\caption{\textbf{UHV test setup and results.} (\textbf{A}) Schematic diagram illustrating the test setup of the vacuum system. (\textbf{B}) Pressure evolution (left axis) during bake-out (temperature on right axis). (\textbf{C}) Recovery of vacuum after periods without active pumping. The pump contains a passive element (non-evaporable getters) and an active ion pump. The data above are the pressures recorded following reactivation of the ion pump after a long period of only passive pumping. Note that the lower limit on the pressure that can be read is $10^{-10}$ mbar, and under-range readings are displayed as 10$^{-10}\,$mbar in both (\textbf{B}) and (\textbf{C}). Errors are approximately $\pm1.5^\circ$ C in temperature and $\pm10\%$ in pressure; the size of the symbols exceeds the size of the error bars.}
\label{vacuum}
\end{figure*}

Temperature dependent emission data taken immediately after the removal of the oxide layer can be plotted as an Arrhenius plot, resulting in a gradient proportional to the activation energy that corresponds within 1\,\% to that of a pure magnesium sample (fig. S4). Therefore the behavior seen in Fig. 3F is consistent with a relatively fast removal of a protective top-surface layer leading to the exposure of a thin, unoxidized, Mg-enriched layer at the sample surface. This layer is then rapidly depleted (regime of fit 1) and Mg emission eventually becomes limited by the rate of diffusion of Mg from the bulk to the surface (regime of fit 2). A formation process for magnesium oxides at high temperatures during the M-PBF build process for AlSi10Mg has been suggested in \cite{Tang2017, Simonelli2015}.

XPS measurements were performed on an untreated surface of the AM material and on a machined surface of the same sample (Fig. 3H). The XPS data confirm the hypothesis of a surface layer with a larger content of magnesium oxides. Fig. 3H (left) shows a peak at 1304.2\,eV corresponding to magnesium oxide compounds such as MgO or Mg(OH)$_{2}$, illustrating  the difference in abundance of Mg in the untreated surface (green line) and the machined surface (blue line). Fig. 3H (right) features two peaks for both surfaces. The low binding energy peak around 71\,eV is attributed to metallic aluminum, while the second peak with a binding energy around 74\,eV is indicative of aluminum oxide or aluminosilicate \cite{Sherwood1998}. The relative abundance of Si is similar on both surfaces (fig.~S5). From the XPS spectra, the composition of the unmachined surface was determined to be 22\% Mg, 69\% Al and 9\% Si, while the machined surface was 1\% Mg, 90\% Al and 9\% Si, consistent with the composition of the powder within the measurement uncertainties (see Materials and Methods). The high concentration of Mg on the untreated surface is remarkable, given that Mg only accounts for 0.3\% of the powder composition (see Materials and Methods), but consistent with the hypothesis developed above. Note that XPS only characterizes the surface up to a depth of 10\,nm. On both surfaces, the Mg component was detected exclusively within a range of binding energies consistent with its oxidized forms, e.g. MgO and Mg(OH)$_2$, rather than pure metal \cite{Haider1975}. By contrast, the Si and Al components were detected as pure and oxidized forms. 

Both approaches confirm the existence of a layer with a high abundance of oxidized Mg at the native surface. The mass spectrometry results also demonstrate that this layer can withstand temperatures up to ~$350^\circ$C and reduces the outgassing of volatile species. For high temperature applications of additively manufactured parts (e.g. for use in scanning tunneling microscopes) the development of methods to enhance the stability of this protective layer can be a promising line of investigation.

\vspace{0.3cm}

\noindent \textbf{Performance testing}

\noindent To test the performance resulting from the design and the surface characteristics, the chamber was built and exposed to post processing as described in the Materials and Methods section. For this, the chamber was assembled into a vacuum system using off-the-shelf stainless-steel viewports and tubes, mounted using standard CF flanges, as shown schematically in Fig. 4A. Bake-out of the system was carried out for a period of 120 hours, during which the temperature did not exceed $120^\circ$C. Once baked, the assembly was pumped using a combined ion / non-evaporable getter (NEG) pump (NEXTorr D$100-5$, SAES Getters) and was found to achieve a pressure in the UHV range. The ion pump current reading indicates a pressure $<1 \times 10^{-10}$ mbar, the lowest pressure that can be read via this method. 
The evolution of temperature and pressure during the bake-out phase is shown in Fig. 4B.

For a portable system, operation without external power supplies is highly desirable -- therefore, the effect of turning off the ion pump for up to 48 hours was tested.  During this time span the chamber was pumped only via the passive NEG elements of the NEXTorr D$100-5$. When the ion pump was reactivated, the initial pressure was measured and its decay back to the under-range reading following ion pump activation was recorded. 
The results of these tests are shown in Fig. 4C. These tests demonstrate that the chamber can maintain pressures in the $10^{-10}$ mbar range for over two hours without active pumping and that even after 48 hours, the pressure remains in the $10^{-9}\,$mbar range. After the switch-on of the pump, the chamber returns to below $10^{-10}$ mbar in less than 6 minutes. The background pressure achieved was more than sufficient to permit the capture of a cloud of cold Rb atoms in a magneto-optical trap. Figure 1B shows a fluorescence image of the trapped atomic cloud.

\section*{Discussion}
In summary, we have demonstrated an additively manufactured vacuum chamber, operating in the pressure range below $10^{-10}\,$mbar and used it to trap a cold cloud of $^{85}$Rb atoms. We described the design steps that lead to a significant reduction in weight while retaining the mechanical stability of the chamber. Analyses of the surface structure shows the existence of a Mg-oxide layer with influence on the outgassing behavior of the AM material.  Introducing AM methods to UHV apparatus has huge potential for reducing the size, weight and material consumption of existing systems and enabling new portable systems with increased functionality. 
These can include the use of advanced lattice structures to increase passive cooling rates or enable external cooling, reduce eddy current generation in response to changes in magnetic field and damp or isolate mechanical vibrations. Another promising line of investigation is the use of AM to produce high surface area elements to enhance the efficiency of passive pumping devices such as Ti sublimation pumps. The results demonstrated here will impact on portable quantum sensors, quantum experiments in space applications and on the wider scientific and industrial community relying on high and ultra-high vacuum systems.


\vspace{0.5cm}

\section*{\label{Methods}Materials and Methods}
\noindent \textbf{Material }

\noindent The pre-alloyed, gas atomized AlSi10Mg powder from TLS Technik GmbH, contains Al 89.8 wt\%, Si 9.7 wt\% and Mg 0.3 wt\% with powder particles in the range $10\,\mu$m to $100\,\mu$m, measured using a Malvern UK Mastersizer 3000.

\noindent \textbf{Manufacture} \\
A Renishaw AM250 laser powder bed fusion machine was used to manufacture all test pieces and the final vacuum chamber. The machine uses a 200\,W Yb fiber laser at 1064 nm wavelength and heats the powder bed to $180^\circ$C during build. The parts were built on an aluminum baseplate using a powder layer thickness of $25\,\mu$m. A hatch spacing of $80\,\mu$m, point distance of $70\,\mu$m and exposure time of $220\,\mu$s were selected, giving an effective scanning speed of 318\,mm\,s$^{-1}$. A chequerboard pattern was used to minimize the accumulation of residual stresses and after each layer, the scan pattern was rotated by 67 degrees. 

\noindent  \textbf{Post processing and mechanical machining}\\
Once processed and removed from the build machine, the parts were stress-relieve heat treated on the build plate for 2 hours at 300$^\circ$C and furnace cooled. The parts were removed from the build plate by wire electro-discharge machining and bead blasted to remove any loosely adhered powder particles. Finally, a T6-like series of heat treatments \cite{Aboulkhair2015} were applied to modify the microstructure of the material. The components were subject to a solution heat treatment for 1 hour at $\sim500^\circ$C followed by a water quench and then aged for 6 hours at $\sim 150^\circ$C.
To ensure a good tolerance at the mating surfaces between the chamber interfaces and the standard off-the-shelf UHV vacuum parts, the interfaces were machined, and knife edges and threaded bolt holes added. The internal chamber surface was left in the as-built condition.

\noindent  \textbf{Knife edge tests}\\
The durability of the knife-edges machined in the AlSi10Mg M-PBF material was tested for repeated opening-closure cycles of a test flange. The performance of the seal was assessed using a leak detector (Pfeiffer Vacuum ASM 340).  During all performed tests (up to 10 cycles) no measurable degradation of the seal was detected and the suitability of the selected post-processing heat treatment was thus confirmed.

\noindent  \textbf{Thread torque tests}\\
The performance of the threads machined into the blind holes was tested against the recommended bolt torque (9.5\,Nm for M4, 16.3\,Nm for M6 from Kurt J. Lesker) using standard stainless steel bolts within a similarly threaded test piece with M4 and M6 threaded holes. A torque wrench was used to apply the recommended bolt torque. No failure or noticeable degradation of the threads was observed.

\noindent  \textbf{Assembly}\\
The chamber was cleaned in three stages before assembly --- first with water and detergent, then with acetone and finally with isopropanol. Following this the chamber was attached to standard conflat vacuum components using standard silver-plated copper gaskets. The bolts used to close the joins and compress the gaskets secure directly into the machined bolt holes in the chamber itself. M6 bolts, tightened with a torque of 16 Nm, were used for the CF40 fittings and M4 bolts, tightened with a torque of 9 Nm, for the CF16. \\
\noindent  \textbf{Bake-out}\\
The assembled system was baked for 120 hours at a maximum temperature of 120$^\circ$C.  In order to avoid the possibility of over-ageing and softening the material around the knife edges, the bake-out temperature was kept well below the ageing temperature. During bake-out, a 300 l/s turbomolecular pump (Pfeiffer HiPace 300) was fitted via the angle valve shown in Fig. 3A.





\section*{Acknowledgments}
We are thankful to Ian Maskery$^{4}$, Dominic Sims$^{1}$ and the Manufacturing Metrology Team for helpful discussions. \\
\noindent \textbf{Funding:} This work was supported by IUK project No.133086 and the EPSRC grants EP/R024111/1 and EP/M013294/1 and by the European Comission grant ErBeStA (no. 800942).\\
\noindent \textbf{Data Availability:} All data necessary to support the conclusions of the article is included in the published material. Additional data can be requested from the authors. \\
\noindent \textbf{Competing Interests:} The authors declare no competing interests.


\end{document}